\documentclass[twocolumn]{aastex63}
\usepackage{lineno,xcolor,hyperref}

\newcommand\msun{$M$\mbox{$_{\normalsize\odot}$}}

\newcommand\mini{$M_{\rm init}$}

\newcommand\mdot{$\dot{M}$}

\newcommand\lbol{$L$\mbox{$_{\rm bol}$}}

\newcommand\teff{$T_{\rm eff}$}

\shorttitle{The age of Wd1}
\shortauthors{Beasor et al.}
\graphicspath{{./}{figures/}}

\begin{document}

\title{The impact of realistic red supergiant mass-loss on stellar evolution}

\email{embeasor@gmail.com}

\author{Emma R. Beasor}\altaffiliation{Hubble Fellow}
\affiliation{NSF's NOIR Lab \\
950 N. Cherry Ave., Tucson, AZ 85721, USA}

\author{Ben Davies}
\affiliation{Astrophysics Research Institute, Liverpool John Moores University \\
146 Brownlow Hill, Liverpool, L3 5RF, UK}
\author{Nathan Smith}
\affiliation{Steward Observatory, University of Arizona \\
933 N. Cherry Ave., Tucson, AZ 85721, USA}




\begin{abstract}
Accurate mass-loss rates are essential for meaningful stellar evolutionary models. For massive single stars with initial masses between 8 - 30\msun the implementation of cool supergiant mass loss in stellar models strongly affects the resulting evolution, and the most commonly used prescription for these cool-star phases is that of de Jager. Recently, we published a new \mdot\ prescription calibrated to RSGs with initial masses between 10 - 25\msun, which unlike previous prescriptions does not over estimate \mdot\ for the most massive stars. Here, we carry out a comparative study to the MESA-MIST models, in which we test the effect of altering mass-loss by recomputing the evolution of stars with masses 12-27\msun\ with the new \mdot-prescription implemented. We show that while the evolutionary tracks in the HR diagram of the stars do not change appreciably, the mass of the H-rich envelope at core-collapse is drastically increased compared to models using the de Jager prescription. This increased envelope mass would have a strong impact on the Type II-P SN lightcurve, and would not allow stars under 30\msun\ to evolve back to the blue and explode as H-poor SN. We also predict that the amount of H-envelope around single stars at explosion should be correlated with initial mass, and we discuss the prospects of using this as a method of determining progenitor masses from supernova light curves.
\end{abstract}



\section{Introduction}
The mass-loss rates (\mdot) of massive single stars above 8\msun\ have long been considered a fundamental influence in stellar evolution, with the potential  to change the end fate of a star by peeling away the hydrogen envelope\footnote{Due to the high fraction of interacting binaries among massive stars \citep{gies1987kinematic,kobulnicky2007new,sana2012binary,kiminki2012updated}, stripping of the H envelope to produce Type Ibc and Type IIb SNe is most often accomplished by binary Roche-lobe overflow \citep[RLOF,][]{claeys2011binary,goetberg2018spectral,sana2012binary,smith11,smith2014mass}. In this paper, however, we focus on single-star models where the mass loss is dominated by winds. } \citep[see reviews by ][]{chiosi1986evolution,heger2003how,langer2012presupernova, smith2014mass,meynet2003stellar}. Stellar evolutionary models incorporate mass-loss by utilizing empirical and analytical \mdot-prescriptions, and ultimately make predictions about which initial masses of stars are expected to end their lives as certain flavours of supernovae (SNe). Importantly, the usefulness of these predictions depends on how accurately the input physics reflects observations of real stars.

For massive stars with initial masses $<$30\msun, winds during the main sequence (MS) are minimal (removing only $\leq$ 0.8M$_{\odot}$), so the only opportunity to significantly impact onward evolution via mass-loss is during the cool supergiant phases (\teff $<$ 10,000K). At present, there is no first-principles model for cool supergiant mass-loss in this region of the Hertzsprung-Russel diagram \citep[although considerable progress is being made, see e.g.][]{kee2021analytic}. Evolutionary models such as Geneva \citep{meynet2000stellar,ekstrom2012grids} and BPASS \citep{eldridge2009spectral} have therefore been forced to choose from a number of empirical prescriptions \citep[e.g.][]{nieuwenhuijzen1990parametrization,van2005empirical,kudritzki1978absolute,reimers1975circumstellar,goldman2017wind,de1988mass}, with \mdot s spanning up to an order of magnitude for a given luminosity \citep[see Fig. 1 within][]{mauron2011mass}. The most commonly used prescription is that of \citet{de1988mass}, but this prescription contains large amounts of internal scatter \citep{mauron2011mass} and has recently been shown to vastly overestimate the total amount of mass lost post-MS for the highest mass objects \citep{beasor2020mass}. Further, the \citeauthor{de1988mass} prescription contains \mdot\ and \lbol\ measurements for only a handful of 15 RSGs. The sample itself is heterogeneous in terms of both mass and metallicity, as well as the method used to determine \mdot, and relies on highly uncertain distances, corresponding to a large source of error in the luminosities. \citet{beasor2016evolution} suggested the scatter in these relations is due to a lack of constraint on the initial masses of the stars used to derive the \mdot-\lbol\ relation\footnote{Evolutionary tracks for different masses can overlap in the RSG phase.  For the same $L$ and $T_{\rm eff}$, a more massive star may have a lower \mdot\ due to a higher surface gravity, for example.}. When measuring \mdot-\lbol\ relations for RSGs in clusters, where the RSGs can be assumed to be the same age, metalliciy and initial mass, the dispersion on the relation is greatly reduced \citep{beasor2016evolution,beasor2018evolution}. Recently, we have combined the \mdot-\lbol\ relations for RSGs in 4 Galactic and LMC clusters of different ages, and derived a new initial mass-dependent \mdot-prescription \citep{beasor2020mass}. This \mdot-prescription is calibrated to RSGs with initial masses between 10 - 25\msun, covering the observed mass range for Type II-P SN progenitors. \footnote{The upper end of this mass limit is particularly important \citep[where the de Jager \mdot-prescription more drastically over-predicts mass-loss, see][]{beasor2020mass} since the upper mass cut-off for Type II-P SN production is debated. Observational studies have suggested that the maximum RSG progenitor mass is $\sim$17\msun \citep{smartt2009death,smartt2015observational}, though the statistical significance of this result may be low \citep[see e.g.][]{davies2018initial,davies2020red, kochanek2020on, davies2020on}.  }

Once a prescription has been selected, further complication can arise in the way in which it is implemented in evolutionary models. For example, the most recent Geneva models \citep{ekstrom2012grids} utilise a combination of both \citet{de1988mass} and \citet{van2005empirical}, both of which been shown to overestimate RSG mass loss \citep{beasor2020mass}. The \citet{van2005empirical} prescription is calibrated using stars with extreme mass loss ($>$10$^{-4}$\msun yr$^{-1}$) that are very rarely seen in unbiased RSG samples, and this prescription therefore is likely not applicable for normal RSG evolution \citep[][Beasor \& Smith in prep]{beasor2016evolution,beasor2018evolution,beasor2020mass}. \citeauthor{ekstrom2012grids} also take the additional step of increasing \mdot\ by a factor of 3 when the luminosity of a star exceeds the Eddington luminosity. While this strategy has been duplicated in other studies \citep[e.g.][]{dorn2020short} there is no observationally motivated reason for this increase in \mdot\ during the RSG phase \citep[e.g.][]{beasor2020mass}. The result of artificially enhancing the RSG mass-loss rates in this way is to force single stars with masses $\geq$20\msun\ back to the blue, where they die as H-poor SNe \citep[e.g.][]{ekstrom2012grids}.

Recently, \citet[][hereafter B20]{beasor2020mass} derived a new \mdot-prescription for RSGs, calibrated to objects with initial masses between 10 - 25\msun. This prescription benefits from a number of improvements upon the dJ88 prescription. Firstly, B20 measured the mid-IR excess\footnote{The warm dust that contributes to the mid-IR excess is sensitive to the last $\sim$100s of years of mass-loss, and so is not an instantaneous measurement.} of RSGs in clusters rather than relying on field stars, providing a sample that is unbiased towards high \mdot\ objects. Secondly, accurate luminosities allow stringent constraints to be placed on both the age of the cluster and therefore the initial masses of the stars \citep[see][]{beasor2019discrepancies}. Further, this new prescription is calibrated from a larger sample of RSGs. In total, \citet{beasor2020mass} measured \mdot\ and \lbol\ for 34 RSGs in 4 different clusters (2 Galactic, 2 LMC), whereas the dJ88 prescription is calibrated using only 15. Finally, whereas previous studies relied on field stars with uncertain distances, \citet{beasor2020mass} use only RSGs in clusters for which distance is more accurately known. For two of the three Galactic clusters, new Gaia \citep{gaiadr2} measurements are available, leading to more precise \lbol\ measurements \citep{davies2019distances}.  

In this current paper, we investigate the effect of implementing the new \mdot-prescription from \citet{beasor2020mass} for the cool supergiant (CSG) phase, using the Modules for Experiments
in Stellar Astrophysics ({\tt MESA}) stellar evolution code \citep{paxton2010modules,paxton2013modules,paxton2015modules}, and from this we make predictions about the progenitors of Type II SN. The paper will be organised as follows: in Section 2 we describe the model grid incorporating the new \mdot-prescription, in Section 3 we briefly describe our results and in Section 4 we discuss the implications of our finding on massive star evolution and SN characterisation.

\section{Method}

\subsection{The model grid}
We compute a new grid of stellar models using {\tt MESA} \citep{paxton2010modules,paxton2013modules,paxton2015modules}. For the input physics, we utilise the inlists for the MIST models from \citet{choi2016mist} using {\tt MESA vr7503}, therefore our work can be considered a comparative study with that of \citeauthor{choi2016mist}  In this work, we consider only the effect of changing mass-loss during the CSG phase, defined as  where \teff $<$ 10,000K (see below for full description), all other parameters remain identical to that of the MIST models. As the B20 \mdot-prescription is calibrated using clusters  where \mini\ is between 10-25\msun, we compute models for stars with initial masses 12, 15, 18, 21 and 24\msun.  This also covers the observationally inferred  mass range for Type II-P SN progenitors \citep[e.g.][]{maund2004massive,fraser2011sn,smartt2015observational}.

In the MIST models, the mass-loss rates for high mass stars (defined as objects with \mini $>$ 10\msun) are implemented using a combination of wind mass-loss prescriptions named {\tt Dutch}. This combination includes \mdot\ prescriptions for hot phases \citep{vink2000new,vink2001mass}, cool phases \citep[][hereafter dJ88]{de1988mass} and another for Wolf-Rayet phases \citep{nugis2000mass}. When stars reach \teff $<$ 10,000K, the {\tt Dutch} recipe switches on the cool phase \mdot, for which the empirically derived prescription of \citet{de1988mass} is utilised, 

\begin{equation}
    \dot{M}_{\rm dJ} = 10^{-8.158}(L/L_\odot) ~ T_{\rm eff}^{-1.676} ~ {\rm M_\odot yr}^{-1}
\end{equation}

\noindent Below, we discuss the new \mdot-prescription from \citet{beasor2020mass}.


\subsection{The updated \mdot-prescription}
 \citet{beasor2020mass} derived an \mdot-prescription that is dependent on the initial mass of the object as well as luminosity, given by,

\begin{eqnarray}
\log(\dot{M}/M_\odot{\rm yr}^{-1} ) & = & a + b (M_{\rm init}/M_{\odot} ) \nonumber \\
&&+c\log(L_{\rm bol}/L_\odot)
\end{eqnarray}

\noindent  where $a = -26.4\pm0.7$, $b = -0.23\pm0.05$ and $c = 4.8\pm 0.6$. Importantly, \citet{beasor2020mass} demonstrated that \mdot\ prescriptions that are based on field stars, e.g. dJ88, vastly over predict the mass-loss particularly for the highest luminosity objects, i.e. the highest initial mass stars. For example, the dJ88 prescription overpredicts the mass-loss of 25\msun\ RSGs by a factor of 40 \citep[see Fig. 4 within ][]{beasor2020mass}\footnote{Though not explored here, it is important to note that mass-loss rates during the hot phases are also thought to be lower than previously derived prescriptions suggest, see \citet{sundqvist2019winds}}. 

We now take the B20 prescription and investigate the effect of lower \mdot\ across the CSG phase on the final fate of massive stars. We compute the models using the same inlists as for the {\tt MESA} MIST models \citep{choi2016mist} until the cool supergiant phase, where the Dutch prescription kicks in. At this point, we switch the mass-loss to the prescription presented in B20, Eqn 2.

\begin{figure}
    \centering
    \includegraphics[width=\columnwidth]{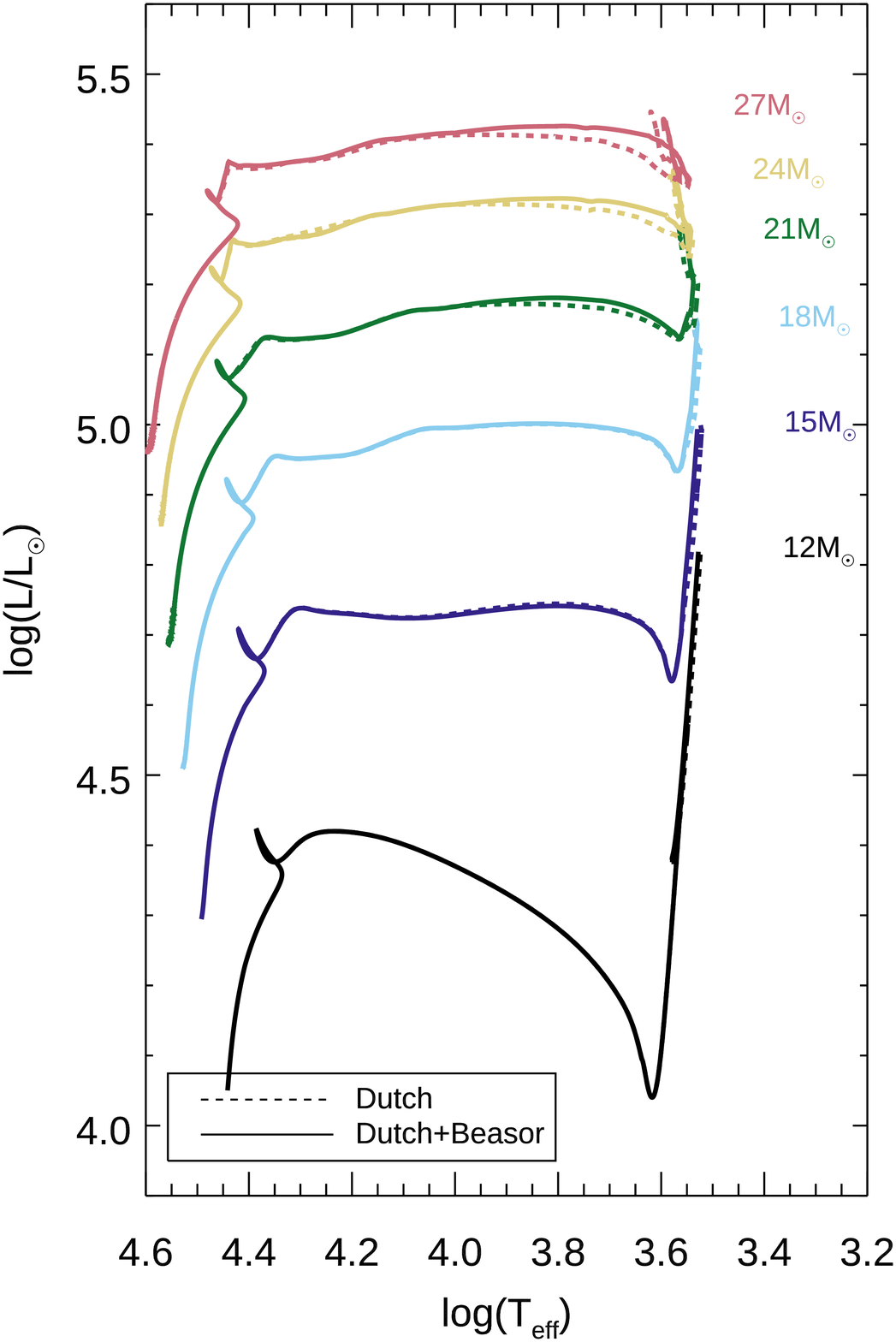}
    \caption{Evolutionary tracks for stars with initial masses using the standard {\tt MESA v7503} \mdot\ treatment (dashed lines) and using the B20 prescription (solid lines). }
    \label{fig:HRd}
\end{figure}

\section{Results and discussion}
In Fig. \ref{fig:HRd} we show the HRDs for each set of models; those with standard MIST parameters, i.e. utilising the Dutch \mdot\ recipe only (dashed tracks); and those where the cool \mdot\ phase uses the B20 prescription (solid tracks). The \mdot-prescriptions switch from the hot star regime to the cool star regime at 10,000K (log$T_{\rm eff}$ = 4.0). As can be seen, the effects of switching the mass-loss recipes has a minimal effect on the location of tracks in the HRD, indeed there are only slight temperature changes for the higher mass stars. \citet{renzo2017systematic} and \citet{sukhbold2018high} also showed that even dramatic downward revisions in CSG mass-loss have minimal effect on the HRD and the final luminosity of RSGs. In the mass range 12-27\msun, there is no noticeable difference between the end point for RSGs between the models employing only the Dutch prescription (dashed lines), and the new models utilising a combination of Dutch and B20 (solid lines). We also do not see any significant difference in the luminosity time evolution of the two model grids.  

Though the final temperatures and luminosities of the stars do not change significantly when the new \mdot\ law is implemented, the terminal envelope mass is substantially altered. The amount of H-envelope at the end of a star's life is an important parameter in determining the rates of various flavours of SNe. For example, should a star retain a large amount of H-envelope (mass) it will appear as a H-rich Type II supernovae, likely a Type II-P. On the other hand, if a star sheds nearly all of its H envelope, it may appear as a Type IIb and the progenitor may appear as a yellow or blue supergiant instead of a RSG. In Fig. \ref{fig:Henv} we show the mass of H-envelope remaining at core carbon burning for the standard \mdot\ implementation (dJ88) and for the new B20 prescription. While the evolutionary paths of the RSGs do not change (i.e. in both model grids the stars would all explode in the RSG phase) the amount of H-envelope remaining changes drastically. 

\subsection{Low mass-loss rates prevent RSGs returning to the blue}
How much hydrogen a star loses during the CSG phases is thought to be a primary driver in the evolution of single stars below 40\msun\ to WRs. For example, in the Geneva models \citep{ekstrom2012grids} a star with an initial mass of $\sim$20\msun\ loses a large enough fraction of its envelope through quiescent mass-loss that the star is forced to evolve back to the blue side of the HR diagram (see earlier discussion). This has been suggested as a single star evolutionary pathway for producing WR stars, and even low-luminosity LBVs in a post-RSG phase \citep[e.g.][]{groh2013massive}. 



\begin{figure}
    \centering
    \includegraphics[width=\columnwidth]{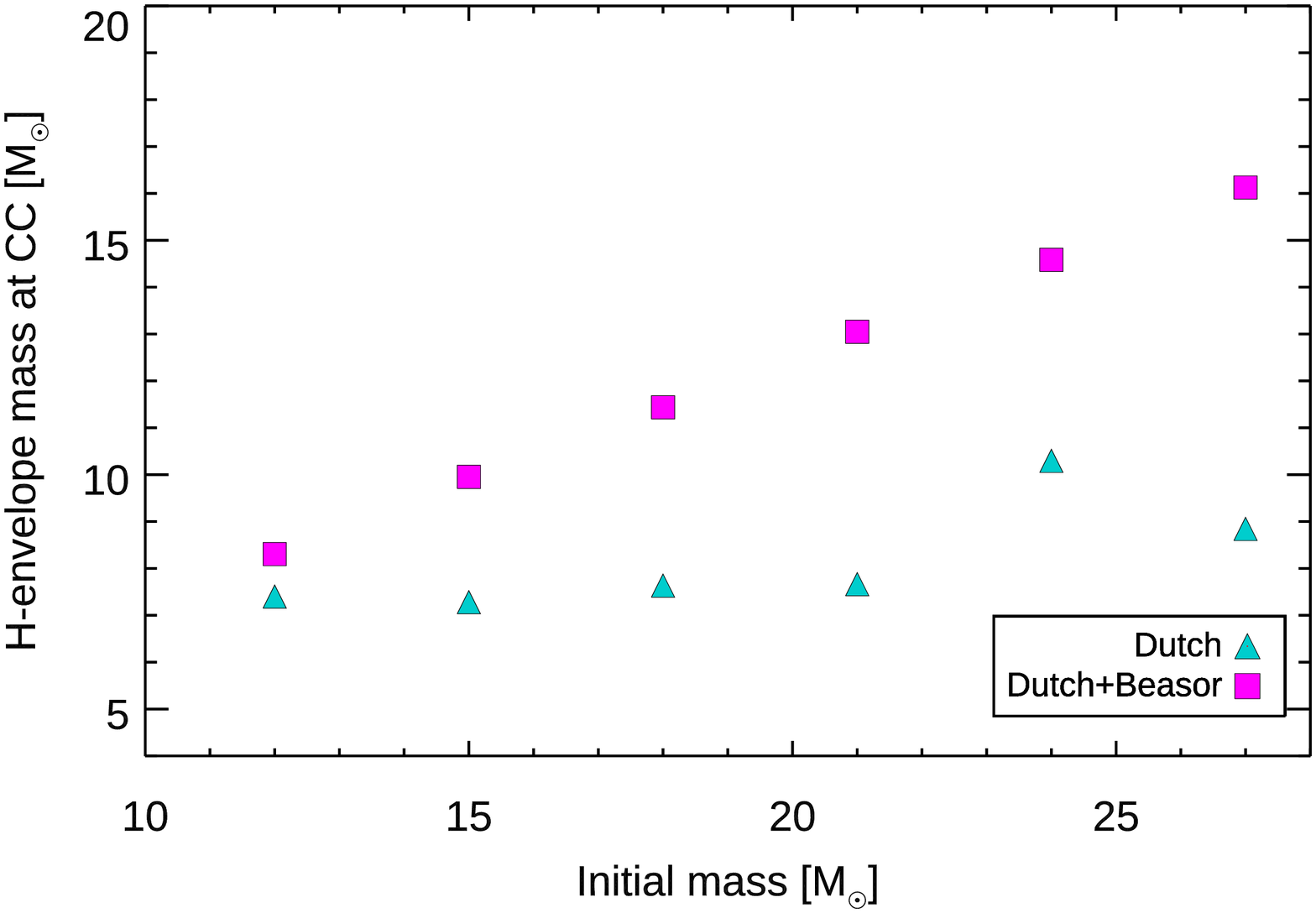} 
    \includegraphics[width=\columnwidth]{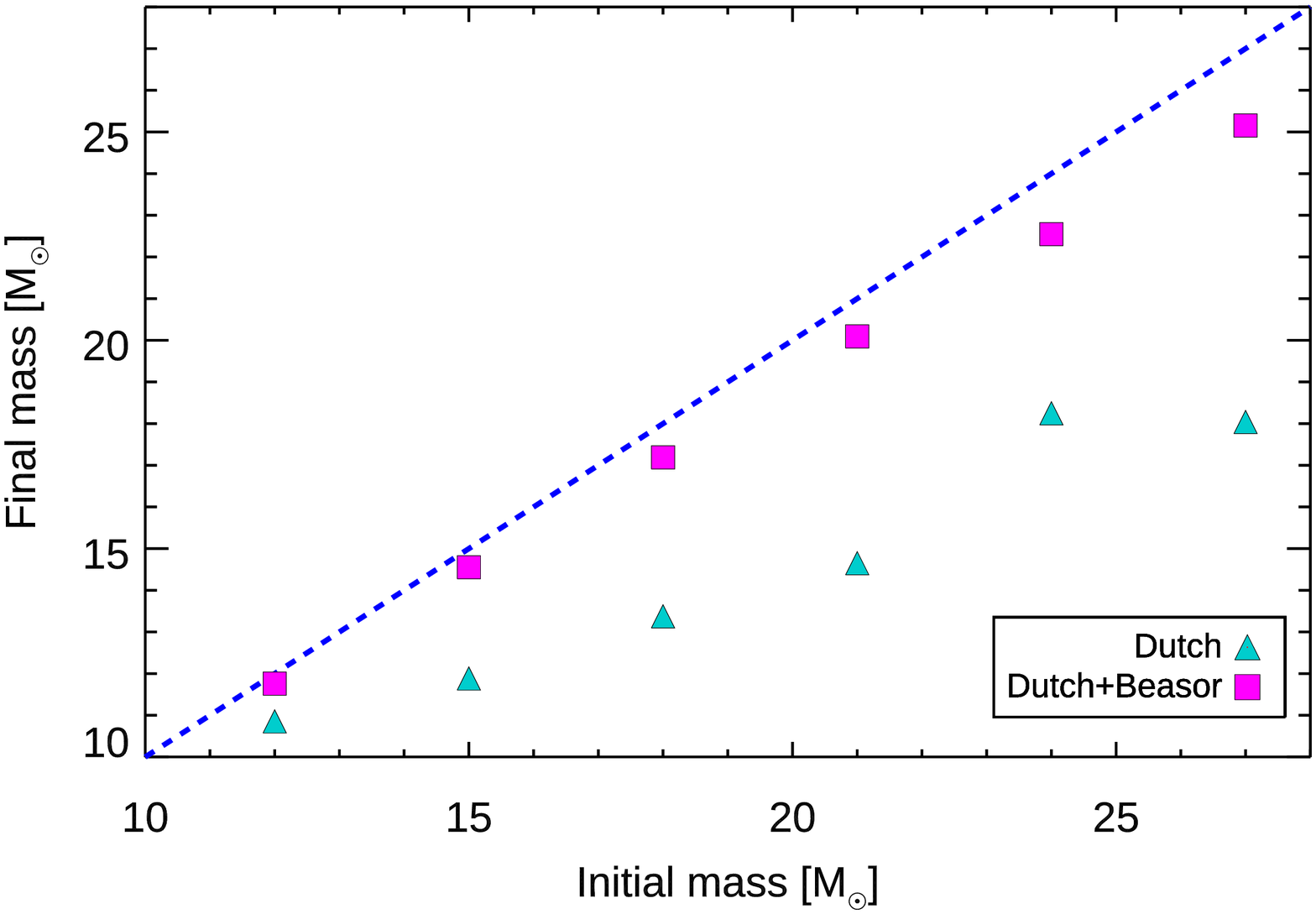}
    \caption{{\it Top:} The amount of H-envelope left at core-collapse as a function of initial mass. {\it Bottom:} The initial-final mass relation for each set of models. }
    \label{fig:Henv}
\end{figure}

Here, we focus only on the MIST models. In Fig. \ref{fig:Henv} we show the amount of H-envelope left shortly before core-collapse (in this case, at the end of core-carbon burning) for both the Dutch prescription models and for models incorporating the B20 prescription. Though the {\tt MIST} models do not artificially enhance the \mdot\ in the same way as the \citet{ekstrom2012grids} models, the figure clearly demonstrates that the amount of H-envelope left at core-collapse is significantly higher when using the lower \mdot s across all initial masses included here. The discrepancy between the two model sets is most apparent for the highest mass objects, i.e. the stars that are missing from progenitor studies, and whose fate is the most debated \citep[see e.g.][]{smartt2009death,davies2018initial,davies2020red,kochanek2020on,davies2020on}. When incorporating either the B20 prescription {\it or} the unaltered dJ88 prescription, the resulting evolutionary calculations show that these high mass objects would not lose enough mass through quiescent winds to evolve to the blue of the HR diagram, and hence would likely die as H-rich RSGs, in contrast to the predictions of \citet{ekstrom2012grids}.\\

Here, our evolutionary calculations predict a large H-envelope mass remaining prior to core-collapse, in contrast with previous models. If correct, it means that steady wind mass loss is not a viable way to make stripped envelope SNe (Types Ibc and IIb) from single stars, and the evolution of stars between 20-30\msun\ to stripped SNe cannot explain the apparent lack of high mass Type IIP SN progenitors.  

Note, however, that the B20 prescription used here does not account for any possible late-phase pre-SN mass loss in the final $\sim$10$^{4}$ years before core-collapse. If such high mass-loss phase were to operate, it is unlikely that the resulting SN would be a SNe Ibc or IIb. Shedding many solar masses of material within such a short timeframe would likely lead to strong CSM interaction, creating a Type IIn event \citep[e.g.][]{smith2009red}. While there is ample evidence for strong  pre-SN episodic mass loss from SNe IIn \citep[for a review, see][]{smith2014mass}, there is currently no prescription for including that mass loss in evolutionary models.

\subsection{Correlation of H-envelope with initial mass}

An interesting outcome of this work is the near-linear correlation at solar metallicity of initial mass with H-envelope mass at core collapse, see Fig. \ref{fig:Henv}. This is the first time such a correlation has been predicted. In the MIST models (which use the dJ88 prescription), the envelope mass at core-collapse appears to plateau, with the H-envelope mass peaking at 10\msun\ for the 24\msun\ model, before declining again for the 27\msun\ model \citep[where the dJ88 prescription most severely overestimates \mdot, see][]{beasor2020mass}. However, when using the B20 prescription, we do not see this leveling off of envelope mass at higher initial masses. Instead, we find a direct correlation between initial mass and envelope mass (though the exact slope may be sensitive to details of stellar evolution that determine the core mass as a function of total mass). 

This raises the question of whether measuring the mass of the envelope at SN could provide another constraint on the initial mass of the progenitor\footnote{The mass of the core may be necessary to determine an initial mass. The core mass can be estimated from the pre-SN luminosity of a star, see relations within \citet{farrell2020snap} and \citet{zapartas2021effect}}. If envelope mass is correlated with initial mass, there may be a signature of this in the SN light curve. Many groups use hydrodynamical modelling of SN light curves to determine initial masses of SN progenitors \citep[e.g.][]{morozova2017unifying,dessart2017explosion,eldridge2018curve,utrobin2017light}, however none of these studies have allowed the envelope mass to be a free parameter\footnote{Some semi-analytic studies have been completed to study the effect of varying parameters on the SN lightcurve, see e.g. \citet{nagy2014semi}.}. 

Hydrodynamical modelling of SN light curves requires input initial conditions, usually taken from the end points of stellar evolution models \footnote{though it is also possible to avoid using stellar models and rely on initial conditons, see \citep[e.g.][]{bersten2011hydro}}. While the precise initial model varies from group to group, in all cases, the progenitor models used as the starting point have all been evolved using the dJ88 prescription during the CSG phases. \citet{dessart2019difficulty} take progenitor models from {\tt MESA} \citep{paxton2010modules,paxton2013modules,paxton2015modules} at initial masses of 12, 15, 20 and 25\msun, including the standard Dutch prescription for CSG mass-loss. The authors find that all models reach the end of the RSG phase with roughly the same envelope mass, 8--9\msun. This means that when these progenitor models are exploded and the resulting light-curve calculated, envelope mass is essentially fixed at $\sim$10\msun\ regardless of initial stellar mass. This raises the question of whether measuring the mass of the envelope at SN {\it could} provide another constraint on the initial mass of the progenitor. 

A high envelope mass at core-collapse may not be in conflict with observations of SN. In SNe II-P, the length of the plateau is controlled by recombination, and hence there may be degeneracy between envelope mass and $^{56}$Ni mass. A higher envelope mass with a lower $^{56}$Ni mass may show the same plateau length as a lower envelope mass with a higher $^{56}$Ni mass\footnote{For example, SN 2009ib had a particularly long plateau phase (130 days), which may be indicative of a higher envelope mass}. We will explore this degeneracy further in future work.

\subsubsection{Impacts of a high H-envelope mass on pulsational instability}
Retaining a high envelope mass may also prevent an RSG from experiencing possible late-phase pulsational instability. RSGs are known to be pulsationally unstable \citep[e.g. ][]{heger1997pulsations}, and it has been suggested that strong winds may lead to a ``superwind'' phase, capable of significantly stripping the H-envelope prior to SN. While these superwinds have not been observed, \citet{yoon2010evolution} investigated the potential impact of a pulsationally driven superwind (PDSW) phase on the appearance of SN progenitors. In this work, the authors found the PDSW kicks in when the $L$/$M$ ratio reaches a critically high value, i.e. when the mass of the star is low. For this to be possible a star would have to lose a significant fraction of its mass prior to a PDSW phase. While \citet{yoon2010evolution} predicted the PDSW phase would be achieved by RSGs where \mini\ $\geq$ 17\msun, they adopted the dJ88 prescription in their stellar models, which (as shown here) overestimates RSG mass-loss. The lower \mdot\ and higher remaining H envelope mass may push the instability threshold to a higher mass, or even prevent high mass RSGs from reaching the critical L/M ratio necessary to drive a PDSW phase.   

If some other mechanism \citep[i.e. wave driving or other energy deposition in late nuclear burning phases;][]{quat2012wave,fuller2017pre, smith2014prep} can trigger instability and episodic mass loss, then there are important implications.  Namely, if lower $\dot{M}$s  for steady winds throughout the RSG phase leave a progenitor with higher H envelope mass, then there is potentially more envelope mass to lose in the final years of the star's life.  This may give rise to more massive CSM shells immediately surrounding the star at the time of core collapse, and hence, more luminous SNe IIn from RSGs because they can experience stronger CSM interaction.

\subsection{Metallicity}
In this work we have only explored the implications of a reduced \mdot\ on Solar metallicity ($Z = Z{\odot}$) models. We note that observational studies of RSG mass-loss suggest \mdot\ is only weakly dependent on Z \citep{goldman2017wind,beasor2020mass}. In addition, at lower metallicities (e.g. LMC and SMC) mass-loss rates are weaker than at Solar $Z$, and therefore would have an even smaller impact on the evolution of an RSG than the Solar models presented here \citep[e.g.][]{mauron2011mass}.

\section{Conclusions}

In this work, we have explored the impact of incorporating the \citet{beasor2020mass} RSG mass-loss rates into stellar evolution calculations. Our main findings are as follows:

\begin{enumerate}
    \item We confirm the results of previous studies, that even vastly reducing mass-loss during the CSG phases has a minimal effect on the final position of the star on the HRD, and the \mini-$L_{\rm fin}$ relation remains unchanged. 
    \item The amount of H-envelope lost due to cool supergiant mass-loss is not enough to cause a star to evolve back to the blue, effectively ruling out the single star pathway for the production of WRs and stripped-envelope SNe at masses below 27\msun.
    \item We find a clear correlation of initial mass with envelope mass at core-collapse for single RSGs. We discuss several implications of the higher H envelope mass for SNe and SN progenitors.
\end{enumerate}

\acknowledgements
We would like to thank the anonymous referee whose comments helped improve the paper. The authors would also like to thank Luc Dessart for useful comments on the manuscript. ERB is supported by NASA through Hubble Fellowship grant HST-HF2-51428 awarded by the Space Telescope Science Institute, which is operated by the Association of Universities for Research in Astronomy, Inc., for NASA, under contract NAS5-26555. This work makes use of the IDL software and astrolib. 

%




\bibliographystyle{aasjournal}
\bibliography{sample63}{}



\end{document}